\documentclass[a4paper, 12pt]{article}
\title{String Theory and General Methodology; \\ a Reciprocal Evaluation\footnote{This article has been published in \emph{Studies in History and Philosophy of Modern Physics}. The published version has the slightly different title ``String theory and general methodology: A mutual evaluation''.} }
\author{Lars-G\"{o}ran Johansson \& Keizo Matsubara}
\date{}
\begin{document}
\maketitle
\begin{center}    
Department of Philosophy\\
Uppsala University\\ 
Box 627, SE-751 26 Uppsala, Sweden \\
{\tt
lars-goran.johansson@filosofi.uu.se, keizo.matsubara@filosofi.uu.se}
\end{center}
\begin{abstract}
String theory has been the dominating research field in theoretical physics during the last decades. Despite the considerable time elapse, no new testable predictions have been derived by string theorists and it is understandable that doubts have been voiced. Some people have argued that it is time to give up since testability is wanting. But the majority has not been convinced and they continue to believe that string theory is the right way to go. This situation is interesting for philosophy of science since it highlights several of our central issues. In this paper we will discuss string theory from a number of different perspectives in general methodology. We will also relate the realism/antirealism debate to the current status of string theory. Our goal is two-fold; both to take a look at string theory from philosophical perspectives and to use string theory as a test case for some philosophical issues.
\end{abstract}
\newpage

\section{Introduction}
String theory has evolved into a dominating field of research, perhaps the dominating one, in fundamental theoretical physics during the last 30 years or so. For a long time the hope among those in the field was that it should be \emph{The Theory of Everything}, that it should give us the answers to all the remaining problems in fundamental physics.\footnote{The world 'everything' should not be taken literally; everything \emph{physical} is the intended scope!} But the hope has so far not been fulfilled and signs of increasing uneasiness  in the  theoretical physics community can be seen. There has always been some opposition against the theory and  in recent years it has gained increased strength. The critics that probably have received most attention are Smolin and Woit. In \emph{The Trouble with Physics} by Smolin (2006) and \emph{Not Even Wrong} by Woit (2006) string theory's dominating place in theoretical physics is questioned. Their main argument is string theory's lack of new testable predictions despite heavy efforts by a huge number of devoted physicists. But the majority of string theorists do not seem deeply concerned; most still seem to be in good mood, although the attack by Woit and Smolin has not passed unnoticed.\footnote{Cf. for example Chalmers (2007).} 

Our first rough assessment of the situation is this: the majority of string theorists are not convinced by the criticism, but they feel a need to come up with a defense. This is profitable for the philosopher, because such a debate will stimulate people to formulate hitherto more or less tacit assumptions and norms. Obviously, those involved do not think the present lack of testability is reason enough to give up. String theorists have not abandoned the quest for testable results, as judged from what they say and write; on the contrary, it is still eagerly wanted. But the situation is problematic and warrants further discussion.  

In this paper we will discuss what philosophy of science can contribute to the debate concerning the scientific status of string theory.  Earlier discussions on string theory have often become too polarized and we believe also somewhat too simplified. Here we will try to give a more nuanced discussion and take a look at string theory from a number of different perspectives on science. We will discuss what conclusions regarding string theory that could be drawn from the perspectives of, respectively, logical positivism, Popper, Kuhn and Lakatos. For a person with a strong allegiance to a specific position in philosophy of science, this might facilitate a decision on how to evaluate string theory.

String theorists hope that their efforts will result in an all embracing theory, a theory that explains everything. Even a superficial reading indicates that the explanation wanted is connected to realistic attitudes, whereas some critics have less metaphysical convictions. Hence the realism/antirealism issue is connected to questions about form of explanation and explanatory value.

An important point  is that one must distinguish between the mathematical framework of string theory and its physical applications. In this text we assume that string theory is a theory attempting to formulate a quantum theory of gravity and unify the fundamental forces, the single exception being when we describe how string theory was originally thought of as describing strong interactions. Applications of the mathematics of  string theory to other areas is not considered to be what we mean by 'string theory'  in this article. 

Another topic that is relevant is whether to adopt an externalist or an internalist perspective on scientific change. Our own views on these questions will be revealed at the end of the paper, but the main purpose of this article is to illustrate how different assumptions regarding science lead to different evaluations of the scientific status of string theory. 

Theoretical physics asks the most profound and general questions about physical nature such as: What are the ultimate constituents of the universe? Is there one unified theory of everything?  How did the universe begin? Is there any explanation of Big Bang?  String theory aims to answer some of these questions.

Such questions are intimately connected to  philosophical issues. For example, what kind of explanation do physicists have in mind? Why is a unified theory of everything desirable? What norms are used when deciding which is the best alternative route to develop an area of research? What needs to be explained and what can be taken for granted, not requiring explanation?

In this paper we will address some of these philosophical issues; we will try to describe and assess the ontological, epistemological and methodological assumptions made by the string theory community. We will also assess the criticism directed towards research in string theory. As has already been stated, we will do so, not by just taking one view on science and scientific method for granted in considering the scientific status of string theory, but to discuss the questions from different perspectives. 
But first we need, as background, an overview of string theory and theoretical physics in general.

\section{String theory - a very brief overview}

String theory has evolved tremendously since its inception around 1970.\footnote{Here we give a very brief description of this development. References to some important articles  are given.  For our purpose it is not necessary to do justice to the finer points of the theory or its historical developments in any detail; only a short overview of string theory's development, an overview that can function as background for the rest of the discussion, is required. To find more detailed descriptions and further references we refer to the vast literature on the subject. Standard textbooks are Green, Schwarz \& Witten (1987), Polchinski (1998) and Zwiebach (2004). Suitable accounts for the layman written by professional researchers in string theory are Greene (1999, 2004) and Susskind (2005). The books by the critics Smolin (2006) and Woit (2006) also contain easily accessible descriptions of string theory and its history. For a discussion on string theory written by a historian of science see Galison (1995).} The basic idea is to explain the multitude of different particles as being different vibration states in quantized strings.  It started off as a theory about the strong interaction. Originally a mathematical formula for certain scattering amplitudes satisfying a number of requirements, which were suggested by experiments and theoretical considerations, was found by Veneziano (1968), but there was no interpretation of this formula. The formula was later, by a number of researchers, derived using the assumption that we were dealing with quantized extended strings instead of point particles.\footnote{ Susskind (1970), Nambu (1970) and Nielsen(1970).} The original version of string theory only contained bosons and needed 26 dimensions to be consistent.\footnote{Lovelace (1971).} Since the real world contains fermions the theory needed to be modified. In this context the idea of supersymmetry was developed and this meant that the theory could deal with both bosons and fermions. It was  then shown that a supersymmetric string theory require 10 dimensions in order to be consistent.\footnote{ Neveu \& Schwarz (1971) and Ramond (1971).}  This is an improvement from the requirement of 26 dimensions; still, it is more than the four dimensions in the observable world and hence an explanation for why the six additional ones are not observed was needed. It was imagined that the extra dimensions were very small and curled up or `compactified'.\footnote{There are other ways of explaining why we do not observe the extra dimension using so called D-branes. This is however of no importance for the arguments presented in this text.}

In the meantime progress was made in conventional quantum field theory to deal with the problems concerning the strong interaction; hence most of those active in string theory abandoned the field and went to quantum field theory instead. This eventually resulted in the formulation of the standard model.

String theory contains a spin two state that could be interpreted as a graviton, the quantum of gravity. This suggested that one could interpret string theory differently, not as a theory for the strong interaction but as an all embracing theory that could unify all fundamental forces and give us a quantum theory of gravity.\footnote{Scherk \& Schwarz (1974)  and  Yoneya (1973).}

However, as long as progress was made towards the formulation of the standard model the interest in string theory was low and only few researchers actually pursued the idea. People also doubted that string theory could provide a quantum theory of gravity free of the  technical problems that had haunted earlier proposals. But string theorists  was able to give convincing arguments for it being indeed possible to formulate a theory of quantum gravity, and after this string theory became increasingly popular.\footnote{Green \& Schwarz (1984).}

Soon it was found that five different supersymmetric string versions could be formulated, viz., type I, type IIA, type IIB, heterotic SO(2) and heterotic $E_8 \times E_8$. The latest view is that these five versions of string theory are connected by dualities to each other and to 11-D supergravity.  They are all thought to be limiting cases of a still more fundamental theory, called `M-theory', still very poorly understood.\footnote{Hull \& Townsend (1995), Townsend (1995) and Witten (1995).}
 
 It was also realized that string theory was not only about 1-dimensional strings. Other many-dimensional objects called branes (a two-dimensional brane is a membrane, thus the generic name) were also accepted to play an integral part of the theory.\footnote{ Polchinski (1995).}

Strong arguments were then given  that string theory allowed an immense number of sufficiently stable vacua.\footnote{Kachru, Kallosh, Linde \& Trivedi (2003).}
This conclusion meant that the hope to derive a unique low energy limit of string theory 
that reproduced the observable phenomena of our world was abandoned by many string theorists. 
Susskind responded by  introducing the term `Landscape' to describe the set of string theory solutions and was one of the early proponents of the use of anthropic arguments in the context of string theory.\footnote{Susskind (2003).}

From this very brief overview it appears that string theory addresses the following  three main problems of theoretical physics. 
\begin{enumerate}

\item
The four fundamental forces in nature are today understood in terms of two different types of theories. The electromagnetic, weak and strong forces of nature are described by quantum field theory in terms of the standard model of particle physics, whereas the gravitational force is described in terms of Einstein's general theory of relativity. 
Both the standard model and the general theory of relativity have been rigorously tested and are now considered to be established parts of our scientific world-view. Hence both these theories are plausibly empirically adequate within very small margins of error.\footnote{While the standard model and the general theory of relativity have been empirically very successful, the situation is not without problems. Straightforward estimates to calculate the cosmological constant based on quantum field theory give results that are way too high, by many orders of magnitude. This is a very interesting problem but will not be discussed further in this article. The interested reader can read more about the problem of the cosmological constant and different suggested solutions in for instance Rugh \& Zinkernagel (2002).}

Unfortunately, the quantum field theory of the standard model is not consistent with the general theory of relativity. This wouldn't cause much trouble for the physicist, in terms of predictions, if we could ignore either quantum effects or gravity in every possible situation. Often this is in fact possible, but not always; there are situations, for instance the conditions that obtain at the very early universe or in black holes, where both quantum effects and general relativity must be taken into account at the same time, and in such situations our ability to make predictions fail; we have no recipe for simultaneously applying the principles of quantum theory and  those of relativity. Moreover, physicists seem to think that it is not satisfying with two unrelated theories about the physical world; there should be only one theory that accounts for everything.
This problem, of finding a unified theory for all the four (known) forces, is the main motivation for string theory.

\item 
Another thing that concerns physicists is the large number of different particles and constants of nature. Their vast number make physicists wonder whether they really are fundamental and to look for a theory that explains them all as manifestations of something more fundamental. The string assumption provides just this. 

\item
A third problem is that the values of the free constants in the standard model are inserted by hand. In recent years this appears less and less satisfying; one wants an explanation. 
\end{enumerate}

For quite some time string theorists hoped that string theory would provide answers to these questions, but in recent years they have become more pessimistic, in particular about the possibility to derive unique values of the constants of nature.  As stated above, it seems that there are many different models of string theory and that our universe is a realization of one of them. What this means for the question of the scientific status of string theory and its empirical testability is something that we will return to later. Now we will discuss how the present status of string theory would be described from the view-point of each of four general methodologies presented by philosophers of science.

\section{String theory and philosophy of science}

 Four well-known theories regarding the development of science are  logical positivism, Popper's falsificationism, Kuhn's theory of scientific revolutions and Lakatos' theory of scientific research programmes.  We will here give very brief descriptions of these views and discuss what an adherent to the respective view would say about the present state of string theory.

\subsection{String theory from a logical positivist's perspective}

The logical positivists view on science have been severely criticized and it is not a popular account of science today. Nevertheless the view has been very influential in past decades and some of the ideas supported by the logical positivists are still influential on how scientists view their theories.

It should be noted that the logical positivists considerably developed their views during its heydays and there was no consensus on all issues, so the presentation here is indeed simplified.\footnote{See Carnap (1966), Nagel (1961) and references therein. A detailed discussion on logical positivism can be found in Suppe (1977).}
The main goal for logical positivists was to eliminate metaphysical discussions from science and secure a firm empirical basis for the sciences. To achieve this goal they proposed a criterion for empirical meaning of any statement, viz., that it should be given in terms of the methods for verifying it. If no such verification method could be given it had no meaning, i.e., no truth value. The main problem is to apply this criterion to theoretical statements. Their solution and final view  was that laws, and theoretical statements in general, lack truth value; only observation statements have a definite truth value and can be verified. Laws was held to be mere instruments for predictions. This of course presupposes a very sharp distinction between observation statements and theoretical statements and this proved to be one of the profound and unsolved problems of logical positivism. 

Logical positivism could be characterized as consisting of verificationism in semantics, inductivism in methodology and an instrumentalist view on theories. Its weakest spot was arguably the distinction between two types of statements, those having truth values and those who have not.

Now, what would a logical positivist say about string theory? Would he repudiate it \emph{tout court} as speculative metaphysics? 

First, the instrumentalist aspect of logical positivism doesn't accord well with string theorist's own views. Since an instrumentalist holds that theoretical statements are tools for predictions,  he has little reason to ask for a unification of the four interactions, which seems to be the ultimate goal for many string theorists. Asking for unification is most naturally interpreted as expressing a realist attitude towards theories, whereas an instrumentalist is satisfied with any set of assumptions, postulates or hypotheses that make correct predictions of observable events. At most, a logical positivist could endorse the search for unification for pragmatic reasons, if that would make calculations and predictions easier. 

Second,  work in string theory is anything but an example of inductivist methodology, it is rather more like suggesting new explanatory hypotheses;  a number of additional assumptions are introduced by which one is able to reproduce quantum gravity and embed quantum field theories of the types that appear in the standard model. But since no new testable predictions hitherto has been made, one cannot say the new  hypotheses have inductive support. According to logical positivism, observations ideally comes first and theory construction is seen as a systematisation of empirical data. Observation and theory construction is supposed to go hand in hand according to logical positivists, which is certainly not the case in string theory.
 
Third, it seems impossible to interpret string theory as encapsulating semantic verificationism. However, this last point is in our view of very little interest, since few nowadays accept the form of verificationism espoused by logical positivists.

Would a logical positivist dismiss string theory as metaphysical speculation? An adherent of the first version of logical positivism, which required complete reduction of theoretical statements to observational statements, would do so, since string theory fails the verification criterion. The defense that string theory is verifiable in principle using large enough accelerators, suggested by an anonymous referee, doesn't sit well with positivism; it is only when we actually can put our theory in direct connection with empirical observations that the theory can be said to fulfill the verification criterion. But an adherent to the later version, where theoretical statements are viewed merely as calculational devices could be less hostile. If string theory would succeed in reproducing all 'low energy data' hitherto accounted for by  GTR and the standard model, it would have the same status as these two theories, viz., tools for calculation. But so long as no \emph{new} testable predictions has been made, this tool is no better than the earlier ones. 

Summarizing, from a logical positivist's point of view, the basic assumptions in string theory is neither true nor false, they are just tools for making predictions about observable phenomena, and since no improvements in this respect have been made, string theory is unsuccessful.  

\subsection{String theory from a Popperian perspective}

To begin, an important aspect of Popper's views of science\footnote{The standard text is Popper (1959).} was to sharply distinguish between context of discovery and context of justification.\footnote{This was however a view that was also endorsed by the logical positivists.} The importance of this distinction is that within the context of discovery there are no methodological rules; the scientist can use whatever means he finds suitable, including metaphysical speculations, when trying to formulate interesting hypotheses. In contrast, Popper had very strict norms for the testing of hypotheses. He endorsed the hypothetico-deductive method, albeit with some idiosyncracies of his own.

Popper's basic norm is that no theory that cannot be falsified may be called scientific. Second, he holds that the scientist should reject a falsified hypothesis and invent a new one. His third norm is that ad hoc assumptions are strictly forbidden. 

Now let's move to string theory; how would a Popperian evaluate its present state? Has string theory been put to a test yet and has it entered a phase of ad hoc excuses, in either the stronger or the weaker sense, that eliminate the testability of the theory?

From a Popperian perspective it may seem ad hoc to say that there are a number of extra dimensions that are compactified. Defenders might say that extra dimensions is not ad hoc assumptions but a core idea in string theory. But, a Popperian may ask 'why compactify precisely six dimensions?' This idea is obviously geared so as to fit our observations of a four-dimensional spacetime.  
 The crucial thing is whether compactification of precisely six dimensions is in principle independently testable, which is the reasonable demarcation for ad hocness. This question is, so far as we can see, presently undecidable and therefore one could not rule out that this assumption could be tested in the future. On the other hand, we cannot dismiss the feeling that the assumption of compactification of exactly 6 (or 7, if we start from  an 11D M-theory) dimensions is ad hoc; why not all 10, or just one, as Feynman asked in an interview (Davies and Brown, 1988, 194).  

Presently, string theory is not testable in the sense that no new testable consequences, not already implied by the standard model or GTR has been derived; this is the basic criticism against it and it is pretty clear that the critics are inspired by Popper's views. 

But people in the field, Susskind for example, are not impressed. Susskind calls the critics `Popperazzi' and continued by quoting Feynman:
\begin{quote} Philosophers say a great deal about what is absolutely necessary for science, and it is always, so far as one can see, rather na\"{i}ve, probably wrong. 
\\
(Susskind (2005, p. 192),  quote taken from Feynman (1963, pp.  2-7). )
\end{quote}
Susskind thus think that criticism inspired by Popper is not to be taken seriously and he believes to have an ally in Feynman on this point. But he has not, for Feynman's view in methodology is more or less text-book Popperianism, as can be seen from the following quote:

\begin{quote}
In general we look for a new law by the following process. First we guess it. Then we compute the consequences of the guess to see what would be implied if this law that we guessed is right. Then we compare the results of the computation to nature, with experiment and experience, compare it directly to see if it works. If it disagrees with experiment it is wrong. In that simple statement is the key to science. It does not make any difference how beautiful your guess is. It does not make any difference how smart you are, who made the guess, or what his name is - if it disagrees with experiment it is wrong. That is all there is to it. 
.....
You can see, of course, that with this method we can attempt to disprove any definite theory. If we have a definite theory, a real guess, from which we can conveniently compute consequences, which can be compared with experiment, then in principle we can get rid of any theory. There is always the possibility of proving any definite theory wrong; but notice that we can never prove it right. Suppose that you invent a good guess, calculate the consequences and discover every time that the consequences you have calculated agree with experiment. The theory is then right? No, it is simply not proved wrong. In the future you could compute a wider range of consequences, there could be a wider range of experiments, and you might then discover that the thing is wrong.
\\
(Feynman (1992)[1965], 156-7)
\end{quote}
It is also known that Feynman was quite skeptical about string theory, as revealed in an interview Mehra made with him (Mehra (1994), 507). 
 
Leaving aside the views of Feynman,  how did  Susskind  defend string theory against  criticism from a Popperian perspective?  His strategy is argument by historical analogy. He gives a number of examples from the history of science in which a hypothesis that once appeared impossible to test now is not only tested but also universally accepted. Susskind concludes: 
\begin{quote}
Good scientific methodology is not an abstract set of rules dictated by philosophers. It is conditioned by, and determined by, the science itself and the scientists who create science. \\(Susskind (2005), 192)
\end{quote}
We agree with Susskind's general remark here, that rules cannot be settled once and for all by armchair philosophy. But it is equally important to notice that Susskind does not question the requirement of testability, only the, in his opinion premature, conclusion that a particular hypothesis is not in principle testable. So in defending string theory against the critics Susskind makes an inductive move; we have seen many cases of theories that once appeared not testable but later has been possible to test, hence we are well advised to wait and hope that also string theory one day will admit testing.

Is it, then, from a Popperian perspective, correct to say that string theory is not acceptable science? This is far from clear. Popper himself held that wildly speculative assumptions, even metaphysical ones, are admissible in science, if they help develop testable hypotheses. Furthermore, he most probably thought that what is required is falsifiability in principle, not that it must be possible to falsify a hypothesis immediately after its formulation, or at any particular moment of time. So a Popperian could accept string theory as  legitimate research while stressing that it has not (yet) produced any testable hypotheses. The crucial thing is, of course, whether it's lacking testable implications really is a momentary or permanent feature. In contrast to the positivists, in principle arguments are more palatable for Popper, the reason being that he did not  tie his views on scientific method to any semantic theory. So presently unfalsifiable claims are considered meaningful and this accords with the view among string theorists themselves. The most reasonable description of string theory from a Popperian perspective is that it is a viable project to work on, despite it being not yet testable. However it is likely that a Popperian would say that although it is a legitimate area of research it has not yet generated a truly scientific theory.  A Popperian would also stress that it is important to be open with this and not claim success prematurely. It seems that Susskind could accept that. 

The critics think, of course, that the time to give up is overdue, while the defenders say it is not. Popper's methodology, interpreted as `in principle falsifiability' cannot decide between them. 

\subsection{String theory from a Kuhnian perspective}

Kuhn was much more descriptive and less normative in his account of science\footnote{Kuhn (1970)} than Popper and the positivists. Furthermore, he adopted a third-person, i.e., externalist perspective in philosophy of science, which entails that  scientific changes not \emph{always} could be explained using the model of rational deliberation as applied to individual or collective decisions in the scientific community. \emph{Sometimes} it is not scientific reasons that ultimately explain theoretical change; instead social and other external causes are invoked. His main reason for this stance is that the conditions for rational theory choice is not always satisfied.
Kuhn describes the development of science as alternations between periods of normal science and periods of revolutionary science. 

Kuhn considers the periods of normal science to be of more importance than what would be suggested by a Popperian account, which focuses on falsification and the refutation of previously held theories. Kuhn described scientific work within a discipline during a normal period as governed by a paradigm consisting of a set of metaphysical assumptions, formalisms, norms and paradigmatic exemplars of successful solutions to scientific problems. Normally these beliefs and habits are hardly articulated. Failures are usually not seen as evidence against the paradigm; they are quarantined for the time being as anomalies. Scientists working within the paradigm assume that these anomalies can be dissolved by modifications of auxiliary assumptions while the basic convictions are left unchanged. According to Kuhn it is these periods of normal science that is the hallmark of science. It is when you take a number of fundamental assumptions for granted and work with `puzzles', which you try to solve in light of these assumptions, that most scientific results are accumulated. If the scientists constantly were debating fundamental questions they would not make any progress. According to this view it is perfectly OK to use auxiliary assumptions to protect the basic assumptions of the theory. But it might happen that anomalies accrue, scientists begin to feel the pressure of many unsolved problems and they start reflecting upon their basic assumptions previously taken for granted. A crisis is on the way, and this may finally result in a replacement of the ruling paradigm for another one. 

Such a change is labeled  `paradigm shift' or `scientific revolution'; it is a change from one period of normal science to another.  Kuhn does not describe this change as the result of a rational choice made by the discipline, since the condition for rational choice, comparability with respect to goal achievement, or commensurability, is not fulfilled. In short, paradigms are, according to Kuhn incommensurable. 

Many philosophers are highly critical of Kuhn for his incommensurability thesis and we join the critics when it is directed to some extreme interpretations of incommensurability, viz., those that entail complete relativism. But we need not take any stance on that matter at this point. Be as it may, let's look upon string theory with Kuhnian glasses and ask ourselves: is the present status of string theory that of a normal science?\footnote{ Audretsch (1981) holds that Kuhn's theory don't fit  theoretical physics, because there are two paradigms in it, a quantum field paradigm and a geometrical paradigm, the latter being the framework for GTR. Each paradigm completely dominates its domain of application. We don't agree with Audretsch' conclusion; For Kuhn, discipline borders are described in terms of paradigms, so if there are two non-competing paradigms, they define two sub-disciplines. But we agree with his point that  quantum gravity may be seen as an effort to unify two paradigms, a situation which is not discussed by Kuhn.We return to this point later.} Is there anything like a string paradigm? Yes, we think so; there are symbolic generalisations, metaphysical assumptions values and exemplars, which upholds a puzzle-solving tradition. In addition, while there are competitors, string theory has a very dominant status, which is a criterion for being normal science. The odd thing is that it acquired this status before it has been able to make any successful empirical predictions.

Our next question is if one could one use Kuhn's theory to decide whether string theory is in crisis and ripe for dismissal?

Our answer is a qualified no. Kuhn's theory was never aimed at giving a methodology with normative force, not even a recipe for determining the status of current paradigms. It is best viewed as an account of history of science, made in retrospect and from an externalist perspective. Its aim is not to give any effective and usable criteria for when a paradigm will crumble, but rather to describe the historical process in retrospect.

Are there more and more anomalies? Is there a feeling of crisis among string theorists? The answer is presumably no. There have been profound unsolved difficulties all along, of which some are solved, some not, and their number has, as far as we can see, not obviously increased. That some critics have voiced their opinions the last years is a sign of increasing dissatisfaction, of course, but the majority doesn't seem deeply disturbed. Thus, so far no revolution is in sight. But one should keep in mind that that could change rapidly and Kuhns' theory doesn't help us by providing any predictions. Kuhn's theory is, to repeat, basically descriptive and historical.

\subsection{String theory from a Lakatosian perspective}

At first sight, one may say that Lakatos tried to steer a middle way between Kuhn's non-rational and non-normative account of paradigm change and Popper's strongly normative falsificationism. In his seminal paper, Lakatos (1970), he introduced what he called the \emph{Methodology of Scientific Research Programmes}, MSRP for short. It contains a methodological norm, or so it seems, while at the same time it takes into account the feature observed by Kuhn that scientists seldom give up a theory or hypothesis when confronting a few unfavorable test outcomes. 

One important difference between Kuhn and Lakatos is that Lakatos held, whereas Kuhn denied,  that research programmes can be rationally compared; they are either degenerative or progressive and the scientific community, if  rational, chooses the progressive over the degenerative one.

Another important difference between Popper and Lakatos is that Lakatos accepted, while Popper denied, that there are crucial experiments that may confirm a theory. These may be conducted when we have a new research programme as an alternative to an old one; Lakatos maintained that in a crucial experiment, one hypothesis is rejected and the competing one is confirmed. He also differed from Popper in holding that a failed test is by itself not sufficient reason to give up; Lakatos stresses that we do not, and should not, reject a research programme if there are no alternatives in sight; better to work on the problems in an unsuccessful programme than to simply give up. 

Lakatos defined a research programme as consisting of a series of theories and characterized by four components: i) the hard core, ii) the protective belt, iii) the positive heuristic and iv) the negative heuristic. The hard core comprises those assumptions that are common to all theories in the series. That which changes from one theory to another belonging to the same programme is the protective belt. Positive and negative heuristic tells the researcher what to do, and what not to do, when pursuing the research programme. 

Lakatos overarching methodological rule for change of research programmes was: replace a degenerative research programme by a progressive one! The distinction between degenerating and progressive research programmes is made in terms of what comes first, theoretical or empirical development. In a progressive programme, theoretical advancements suggest new experiments that then are planned and conducted. Then the empirical results come out most often as predicted. This is to be contrasted with a degenerative programme where empirical findings force the scientists to modify their theory into newer versions within the programme by replacing components in the protective belt. In a degenerating research programme the function of the protective belt is akin to that of ad hoc assumptions; the difference is that independent testability might be possible. So the methodological rule to give up a degenerative research programme has an effect analogous to Popper's rule not to accept ad hoc assumptions.  

But when has a research programme entered a phase of degeneration so it is time to give up? Even a very progressive programme could come into a degenerative phase for some time, and scientists were by Lakatos advised not to give up too hastily. It is often rational to wait for some time before giving up. In fact, Lakatos has not given us any effective methodological rule by which we could arbiter between rational and irrational actions. Both to give up and to stick to an ongoing research programme could be defended as rational, given Lakatos' prescriptions. 
Unfortunately, Lakatos died before he had answered this criticism, so we don't know what he thought about it. Obviously, Lakatos theory needs improvement.

Or does it? Perhaps we only need a new perspective to evade the criticism? This was Hacking's point of departure in his assessment of Lakatos' theory in Hacking (1983). Hacking's defense was that Lakatos could be reconstructed so as not to give methodological prescriptions for the individual scientist, but to give a method for rational reconstructions of periods of history of science. 
The development of a science during a period can be judged rational or not rational by the standards chosen, post hoc, by the philosopher or historian of science, independently of the beliefs of those working in the evaluated discipline.  
 
Thus, the advice to give up a degenerative research programme in favour of a progressive one might not be an advice directed to scientists but an advice directed to the historian aiming at rational reconstruction of science history. Every historian has to choose a perspective and some principles for selecting which facts to take into account and which to neglect. Thus Lakatos methodological rule is better stated as: reconstruct periods of history of science in such as way that scientists are depicted as normally giving up degenerative programmes in favour of  progressive ones when the objective conditions are fulfilled, disregarding their beliefs. 

It does not follow, from adopting Hacking's interpretation of MSRP, that the concerned scientists, the agents being described, are not given any methodological rules. Knowing about MSRP in Hacking's interpretation, accepting it and wanting to be rational, they should certainly follow the replacement rule if possible. But they have not enough information to definitely judge their alternatives, so the replacement rule is not an effective one for them. It is only effective for the historian.

According to a Lakatosian view one may also accept that many research programmes can exist in parallel and adherents to different programmes may very well agree that one programme is more progressive than the other  at any particular point of time. But there is nothing irrational to pursue a less progressive programme, although it appears that the risk for failure is higher. One has no argument for saying that minimizing risks in chosing research programme is the most rational option. 

If one research programme is particularly progressive it will presumably attract most researchers in the field and other programmes are simply abandoned. This is natural if we believe that scientists share a minimum set of values, attitudes and vaguely defined norms.

After these interpretative considerations it is time to take a look at string theory as an instantiation of a Lakatosian research programme. 
Here is a selection as to what could be a plausible reconstruction of string theory's content, or part of it, in terms of the four components of a Lakatosian Scientific Research Programme :

\begin{itemize}
\item
Hard core: i)The fundamental objects are  not point particles but extended objects (strings or branes).
ii) Accept the basic assumptions of quantum mechanics as given.
iii) Require supersymmetry of the theory.\footnote{We assume that we talk  about superstrings here. It should also be noted that supersymmetry should be broken at lower energies in order to give a correct description of our world.}
\item
Protective belt: 
i) The different versions of string theory are merely different theory formulations, not different theories.
ii) Compactified dimensions are too small to be observed with present day technology.
iii)Explain the value of the constants of nature assuming a landscape of universes. 
\item
Positive heuristic: Develop the theory so one can i) explain the diversity of particles as mere manifestations of one fundamental kind of objects, ii) derive the constants of nature, iii) unite the standard model with gravitation.
\item
Negative heuristic: don't allow any modus tollens argument to be directed against the hard core.
\end{itemize}
Comparing string theory with its competitors, e.g.  loop quantum gravity, it seems clear that  the string programme has attracted most researchers, which have developed the theory in many respects, and so one could say it has been progressive in a more general sense.\footnote{It should be noted that quantum loop gravity while being a contender when it comes to quantizing gravity do not attempt to unify all forces. Despite this less ambitious goal we think it is fair to describe quantum loop gravity as a competitor to string theory.} On the other hand, using Lakatos criteria for distinguishing between progressive/degenerative ones, no such divisions can be made, simply because empirical tests has been lacking.\footnote{For a more detailed discussion on various approaches to the problems of quantum gravity written for philosophers of physics we refer the interested reader to the following sources: Rickles', ``Quantum Gravity: A Primer for Philosophers'' in Rickles et al. (2008), Rovelli's, ``Quantum Gravity'' in Butterfield and Earman (2007), Rickles, French and Saatsi (2006) and Callender and Huggett (2001).} 

However, there is a sense in which string theory confronts the observable world. Take for example constants of nature. We have known their values for some time. String theorists during one period hoped to be able to derive them from first principles. The derivation of the actual constants failed, one got a vast number of possible values and combinations. In response some came up with the idea of a multiverse, consisting of an enormous set of universes, each characterized by a unique combination of values of those constants. This move fits rather well with Lakatos' description of a degenerative phase in which empirical findings drive the theoretical development, although in this case the empirical results were known in advance. 

Is the difference between knowing the empirical results in advance or not relevant for the question whether the programme is degenerative? Of course it is;. in a truly progressive programme \emph{new} empirical results are predicted. But conflicts with previous experiments could still be considered to have some relevance. The move of postulating a multitude of universes as a response to a failed prediction/retrodiction could plausibly be described as an instance of a degenerative phase of the research programme. For there is nothing in the hard core of the research programme indicating a multitude of universes, nor that that assumption is needed for protecting the hard core from being targeted by a modus tollens argument. This does of course not mean that the hypothesis that there are multiple universes necessarily is wrong, but one should at least be a bit worried. The existence of these multiple universes has also been debated among string theorists themselves so it is clear that this hypothesis is controversial even in the field.

 String theory is a degenerative programme, according to Lakatos' criterion; the empirical facts against which string theory is tested was known in advance,  no new testable empirical predictions have been made, and the mismatches that have been found have been a driving force in its development. Hence, if we read Lakatos as proposing a decision rule, the conditions for dismissing a research programme would be fulfilled, \emph{if} there was a progressive rival programme. This is however not the case so we can not really apply Lakatos' criteria and recommend scientists to abandon research in string theory. 

\section {Further remarks concerning string theory and scientific norms}

There is no general agreement about which perspective to adopt in general methodology; hence, the evaluation of the scientific status of string theory must be conditional on which stance one takes. Nevertheless, some general conclusions are possible. The first is that, except for adherents to a strict verificationism typical of the early logical positivists, no matter which of the discussed positions you take, you cannot definitely and clearly reject string theory as unscientific. Second, again with the exception of the verificationist, no methodologist could definitely tell string theorists that it is time to give up and do something else. Perhaps somewhat surprisingly, not even a convinced Popperian could say that string theory should \emph{now} be rejected as a falsified or unfalsifiable theory.

In our own view, the failure of now making a definite judgment about what is rational vis a vis string theory is not at all astonishing. For if a methodologist were able to do that, he would in this particular situation have solved the induction problem; he would be able to tell us, on the available evidence from history of science, what is the correct inference, i.e., he would be able to tell us what is the rational way to proceed. But we don't think that is possible; as Quine once wrote `The Humean predicament is the human predicament' (1969, 72).

We hold Lakatos theory, MSRP, to be the most reasonable analysis of scientific development; it fits quite a number of episodes from history of science. It is also, to some extent, useful for discussing string theory and its competitors, mainly  loop quantum gravity. However one cannot really say that one programme is progressive and one degenerative,  because the distinction and comparison  is made in terms of theoretical and empirical development, and no empirical development has occurred. On the other hand, without using Lakatos criteria and instead merely relying on our somewhat vague notion of development, one is tempted to say that string theory has been \emph{theoretically progressive}, but not \emph{empirically progressive}. One could say that adherents to string theory believe that theoretical progressiveness is sufficient for continuing work on the theory, whereas critics think it's not. 

In a short article Cartwright and Frigg (2007) arrive at conclusions somewhat similar to the ones here presented. The authors end up by evaluating string theory in terms of Lakatosian research programmes, although they have used a broader set of criteria for determining string theory's degree of progressiveness than Lakatos did. They mention such virtues as i) a large range of empirical applications, ii)successful novel predictions, iii) spawning new technologies, iv) answering perplexing problems, v) consistency, vi) elegance, vii) explanatory power, viii) unifying power, and ix) truth. They conclude that string theory has been progressive along the dimensions of explanatory and unifying power, but that is not sufficient for saying that string theory  is generally progressive. The authors however does not recommend dismissing string theory; referring to Lakatos, who is said to recommend us to ``treat  budding programmes leniently" they conclude that string theory deserves still being pursued.  We agree that Lakatos MSRP does not give a definite recommendation in a case like the present one. We also agree that string theory deserves still being pursued, albeit we see no good reason why most efforts in this area should be put on improving string theory; it isn't obviously superior to its competitors. 

Their list of dimensions of progressiveness is hardly an interpretation of Lakatos' notion of progress, it is their own version. Two of their criteria, elegance and truth, deserve a comment. As regards elegance, it could reasonably be held that elegance, however it is made precise, is not a property of the object but  is in the eye of the beholder; it has to do with scientific taste' and other non-epistemic aspects and we don't think such criteria should be used. Second, we question their mention of truth as one dimension along which a research programme should be evaluated. We may infer that a certain theory plausibly is true, or approximately so, using all available evidence. Thus truth of a theory is not a criterion but the aim. 

One important aspect of Lakatos  MSRP is not stressed in Cartwright and Frigg's article, viz., that one should abandon a degenerative research programme in favour of a progressive one. In quantum gravity there are no truly empirically progressive competitors to string theory, hence this rule can not be applied. So what is the suitable strategy to use when no programme is progressive? Neither Cartwright and Frigg, nor Lakatos give a definite answer;  some people may take the risk on working with alternative research programmes before they reach the progressive state, some other may not, and neither choice seems irrational. We return to this point at the end of the paper. 

We think that Kuhn has widely exaggerated  the incommensurability between successive paradigms, at least when applied to science after the scientific revolution. Kuhn did not differentiate between the case where an old theory is believed to be strictly speaking false but still respected as a useful approximation and when a theory is abandoned as wrong tout court.  For example, Newton's theory is still considered good science even though we now hold it strictly speaking not correct, as a contrast to Aristotle's mechanics, which is false and also considered to be useless in scientific practice. The change from Aristotelian to Newtonian mechanics is thus, in a certain sense, much more profound, than the change from Newtonian mechanics to relativity theory and, in the field of micro physics, the advent of quantum theory. 
Furthermore, we do not believe that scientists working in different fields or research programmes have completely different norms; on the contrary, they share several norms.

In Dawid (2009) it is argued that the conflicting assessments of string theory is due to a difference in Kuhnian paradigms. We find these arguments exaggerated. We do not believe that there is a radical difference in methodology or criteria for evaluating success between defenders of different research programmes in quantum gravity and traditional physicists. All involved agree about the need for empirical testing and there is no sign of there being radically different interpretations of empirical observations, which was a characteristic part of Kuhn's argument for there being incommensurable paradigms. Dawid talks about  a `metaparadigmatic shift' since even the methods have changed and not just the view of the world. This talk of `metaparadigmatic' is not needed since  methodological norms are included in the concept of a paradigm  and Kuhn held that methodological norms could differ between different paradigms.

\subsection{Why are tests of string theory required?}

String theory has not, so far, predicted any new observable phenomenon and it is understandable that people lose patience. On the one hand, this criticism is serious; a theory that can't predict anything new appears to be no better than pseudoscience. But on the other hand, string theory's basic aim is not just to make novel predictions but to explain hitherto unexplained facts, so why would lack of predictions problematic? Has the critics misunderstood the goal of string theory?

We think not. We need more than consistency and explanatory power as criteria for selecting the correct explanation, since explanatory power is much too vague;  there is no agreement about what an explanation is, not even how to compare two explanations and telling which is the better one. 

Most theoretical physicists do not seem deeply concerned about the \emph{present} lack of testability. Of course, they all would like a novel testable prediction coming out true, but the lack of testable consequences have not, at least so far, convinced the majority of those doing string theory to give up. Why not? We don't think that they have dismissed testability as a necessary criterion for empirical science, quite the contrary. String theorists simply hope to be able in the future to derive testable consequences.\footnote{For an example of a text where a physicist discusses the need for string theory to confront empirical tests see Schnitzer (2003).}

Some critics might say that instead of looking for reasons given by the researchers themselves we had better look for external factors that in fact \emph{cause} them to continue work within it, causes that do not provide scientific reasons. Looking for reasons is to adopt an internalist perspective, asking for causes that are not reasons is to adopt the externalist perspective. 

\subsection{String theory from an internalist perspective: explanations}

The three prime goals for pursuing string theory are, to repeat, i) to unify all four forces of nature, ii) to explain the vast particle diversity, iii) to derive the constants of nature from first principles. Fulfillment of all these goals, not only the second one, could plausibly be called explanations, so the reasons for doing string theory can be formulated as quests for explanations. This motivates a look at what kind of explanation is suitable in fundamental physics.

It is clear that the hope for an explanation embracing all four fundamental interactions (a `Theory Of Everything') is strong impetus for those doing research in string theory. But how does a satisfactory explanation look like? The answer depends on ones metaphysical position; if your inclination is pragmatist or instrumentalist, holding that all we want of science are reliable predictions, the difference between explanation and prediction is only a difference concerning timing. We predict what will happen and explain what has happened, but this is only a pragmatic difference; the logical structure is the same, often held to be captured by the Deductive-Nomological model of explanation, or its statistical counterpart.\footnote {An explanation is, according to the DN-model a derivation of the explanandum from laws and other assumptions. In the statistical version the laws are statistical.} 

But if you hold that all scientific propositions are descriptions of the world that are true or false, then explanations, in some sense stronger than mere true predictions/retrodictions, is eagerly wanted. Most people are realists in this minimal sense and so for example they might ask themselves: is the correct explanation of particle diversity the one in terms of strings, i.e., one-dimensional physical objects?

\subsubsection{Explanation as unification}

A possible model of the explanatory structure of string theory, and of any fundamental theory, is unification. There are at least two explications of the notion of unification, Friedman (1974) and Kitcher (1981). However neither of these are entirely satisfactory. Albeit it's shortcomings, we hold Friedman's attempt the better one. It's  general idea is easy to grasp: a statement, or a set of statements, are explanatory and functions as explanans if it (they) enable derivations of several laws and empirical regularities that previously seemed unrelated and already accepted as being true.

It seems plausible that the prospect for such unificatory power is the basic reason why many theoretical physicists are so fond of string theory. The basic idea of string theory, that the fundamental objects are one- or multidimensional objects whose internal vibrations explain their observable properties carries great promises for unification. 

The first aspect of string theory's unificatory power is its ability to bring in gravitation into a quantum theory of the world. This tells very much in its favour, in particular since it was unintended. This reasoning has the form of a `theoretical no miracle argument': it would be a miracle that string theory is able to unify the standard model and gravity if it is not substantial truth in it.  In contrast to the postulate of extended fundamental objects however, alternative ways of unifying gravitation and the standard model is not a priori excluded. 

Since GTR was known in advance, one cannot literally say that string theory `predicts' it; it is rather a case of retrodiction. But this is precisely a consequence of having proposed a unifying assumption.  
 
The situation is analogous to that of GTR at its initial stage: GTR entails that the perihelion of Mercury will precess. This precession had been observed since long but no satisfactory explanation was available before GTR.  Since GTR was constructed from a few basic principles, its ability to explain a known  but hitherto unexplained fact count strongly for its at least approximate truth.  

\subsubsection {Explanation of diversity of objects}

Lee Smolin, albeit a critic of string theory, frankly acknowledges that string theory explains the diversity of matter and force particles. The explanation has the form of unification, in a somewhat vague sense. In short, assuming that everything is made up of strings that propagate according to one simple law is all we need to get started. He writes:
\begin{quote}
Indeed, the whole set of equations describing the propagation and interactions of the forces and particles has been derived from the simple condition that a string propagates so as to take up the least area in spacetime. The beautiful simplicity of this is what excited us originally and what has kept many people so excited; a single kind of entity satisfying a simple law. \\ (Smolin (2006, 184).)
\end{quote}

It is easy to feel the pull of such a forceful explanation, and we think Smolin is right in saying that this beautiful simplicity, together with its using previously well-known physical principles, is what makes people continue to hold on to the string programme. 

Let us only point out the epistemological structure of this explanation. It is not the case that we knew in advance that there are strings that propagate according to a certain law, and then physicists were able to show that many phenomena thus could be explained. The epistemic order is rather this: we have a problem, a confusing diversity of different kinds of objects, each kind following specific interaction laws. Then someone suggests two hypotheses: i) all objects are strings, ii) these objects propagate so that the covered area is minimized. Then, using the usual method of quantisation, we can derive the formerly unrelated phenomena. These derivations provide good reason to accept the two hypotheses as true. If so, we have an explanation in the form of unification. 

The strength of the explanation appears to be strongly related to how diversified the things explained are; the more seemingly unrelated phenomena that can be derived from these two assumptions, the more probable these assumptions appear to be, and the better unification we have.

\subsubsection{Explanation of constants}

From the basic theories in physics, the standard model and GTR, one cannot derive the value of the constants of nature; rather, their values seem to be completely accidental, and very lucky ones, because only a minute change in any of them would make it impossible for any form of life known to us. In short, it seems to be a miracle that the physical conditions for life in our world are fulfilled. From this perspective a new and demanding explanation request come to the fore: how do we explain the values of those constants?
 
Since the original hope of being able to derive the unique and correct values of these constants from first principles had to be abandoned, the present idea is to say that different combinations of parameter values exist and correspond to different universes. 

That means that our world is just one of staggering number of worlds, the number $10^{500}$ has been mentioned, Susskind (2005,  290). These worlds are not supposed to be just mere possibilia, but as really existing! Of all these existing worlds, we live in one, and this is explained by the fact that of all possible combinations of values of constants, only one ( or very few) permits life as we know it. This is the anthropic argument. Many physicists feel quite uneasy about it; is this really a viable explanation? 

Our view is that it's validity rests on holding all those worlds as really existing; if they were mere possibilities, the argument would be a teleological one. For suppose we take all the worlds as mere possibilia. Then the first step in the answer to the question, `why do the constants have precisely the values they have?' would be that only our world is actualized. Why is that so? This is explained by saying that this is so because we live in it. This form of explanation would do if we believed that our existence is the result of purposeful actions. But most physicists don't accept such an argument as a physical explanation. We agree.

On the other hand, if we hold that all the different worlds really are existing, the explanatory structure is different. In this perspective no purposeful actions are assumed and different worlds with different possible combinations are realized. Then life evolves, as a matter of course without any intervention, in those worlds where conditions are suitable, and our world happens to be one such. In short, the anthropic argument makes physical sense only if it is taken for granted that many possible worlds are really existing. However, one could then argue that assuming the real existence of the all these worlds reduces the explanatory value because now we have more assumptions in explanans.  

Summarizing, string theory brings about unification all four forces of nature and of the diversity of particles. But the third explanatory demand, viz., explaining the constants of nature is more problematic.
\subsubsection{Other applications of the mathematics of string theory}

 In addition to the demand for unification of fundamental physics, research in string theory has also been justified by its contribution to the development of pure mathematics, in the sense that it suggests interesting conjectures fit for being proved more rigorously. These things are relevant for justifying continued work on string theory, viewed as \emph{uninterpreted mathematics}, but provide no relevant argument for or against the correctness of string theory as a theory about the fundamental interactions. 

A similar remark can be made concerning attempted applications of the formalism of string theory to other topics than fundamental strings and quantum gravity. For example, using the mathematical results behind the understanding of  dualities, attempts have been made to apply the formalism of string theory to other things.  Dualities are seemingly different descriptions and formalisms that have physically equivalent content. For instance a theory of strings can be physically equivalent to a theory of particles and a type of string theory on one background can be physically equivalent to a different kind of string theory on another background and so forth. Exactly what this means and how this should best be interpreted is a complicated and interesting question that we will not discuss in this paper.
One example is the use of string formalism for describing  quark-gluon plasma; see for instance the review by Gubser (2009) and references therein. If such attempts would be empirically successful, it would show the usefulness of the \emph{mathematics} of string theory; but it would not say anything about the string theory interpreted as a theory about the fundamental interactions or its ability to solve the three fundamental problems stated in section 2.

\subsection{String theory from an externalist perspective}

Now let's have a look at string theory from a a perspective in which scientific activities are not viewed as purposeful deliberations and actions but as effects of causes described in a non-intentional idiom.  

In the debate concerning string theory, some critics have suggested that non-scientific factors must be taken into account when trying to understand why people have not left the business, despite its lack of empirical success. Assuming that empirical testability is a \emph{sine qua non} for empirical science,  there must be something else, i.e., factors external to the science per se that has made people willing to pursue this research. 

Another argument put forth by for instance Smolin (2006) is that now almost all important positions in theoretical physics at major universities are held by string theorists. Students are forced into string theory due to lack of alternatives. It is claimed that other approaches to deal with the problems of quantum gravity is not given enough funding and that this is not fair. Is this a telling explanation?

As is well known, string theory was for quite some time pursued only by a few number of individuals. They had little funding, being far beside the focus of interest in theoretical physics. But they came up with results that interested a wider audience and within a short time string theory became mainstream. Today other approaches to quantum gravity have no less funding than string theory had at the beginning. There is reason to think that if people in a programme competing with string theory would come up with good enough results, their approach would most plausibly attract more researchers and get better funding. 

The crucial question is, of course, who is to judge what to count as a `good result'; do adherents to different approaches share the same, or sufficiently similar, norms regarding what to count as progress, or do they differ? If they have sufficiently similar norms there is hope for a fair game. We think that that is in fact the case.

To make the argument by Smolin more convincing one needs to make it plausible that string theorists of today are more stubborn than the previous generation of particle physicists who turned from their previous work into string theory. One possible argument for that can be put like this. The work done in particle physics using the framework of QFT produced many testable results and led to successes.  When scientists previously working in this field turned to string theory they did not thereby nullify the worth of their earlier work. In contrast, if a string theorist would give up and move to another area of research he would thereby publicly show that he thought string theory was a failure. Such a move is plausibly more difficult to take if ones whole career has been confined to string theory. 

However, as we have mentioned above, we do not believe that string theorists have abandoned the idea that experiment is the final arbiter of theories; there is still in the physics community a much higher valuation of  empirically testable and confirmed predictions than merely theoretical results, however interesting.

\section{Conclusions}

We have in this paper discussed the present status of string theory from different methodological views, but also elucidated these methodological views in the light of an ongoing research programme, about which we don't know whether it will be a success story or a dead end.

The long debate in general methodology has not produced anything like consensus concerning norms for rational choice between competing theories/paradigms/research programmes in the same field. This negative outcome is in our view to be expected and to be welcomed. This is so because if there were consensus about a set of universal and permanent methodological norms telling us when to keep on and when to reject a research programme, these could only have been motivated by a priori arguments and independent of actual varying practices. That would mean that philosophers of science would have agreed upon a kind of first philosophy, an epistemology  independent of empirical and thus revisable considerations. We believe such a position is untenable. We wholeheartedly accept that epistemology is part of empirical science and is open for revision, as all empirical science. In short, we accept the basic tenet of epistemological naturalism. 

It is common to argue for a particular methodology by using  success stories from history of science, which by their historical nature are such that we know the outcome and  we know that the scientists  made the 'right' decisions. It has, for example, since long been agreed that the atomic theory of matter is true and those who opted for it took the correct decision; they were rational.  So one can raise support for a particular methodology by describing this theory such that it fits the general methodology. But the force of the example comes from the success.   Looking at an ongoing research programme/paradgim/theory such as string theory whose fate is unknown,  shows us that none of the well-known views in general methodology gives  a clear recommendation on how to continue.

An anonymous referee describes us as concluding  that 'the jury is out' and wrote that that is old news. Thus he/she implicitly seems to want a stronger normative stance on string theory. But  we reject the idea that philosophers of science can act as jury members in telling scientists whether a research program should be given up or not.  Philosophers have no such privileged stance from which to decide such things. The role of philosophers of science is to be active participants in an ongoing discussion on science and scientific methods but not to lay down the law once and for all.

There is little reason to claim that theoretical physics in general, and string theory in particular, has lost its aim predict outcomes of new experiments. String theorists would of course consider the derivation of a testable prediction as big success and if the prediction would come out true, the victory would be telling. That string theory so far has not been able to come up with any testable consequences is disappointing, but not reason to lessen the demands on physical theories and string theorists themselves shows little sign of changing their mind on this issue.

Hedrich (2006) has a somewhat different view; he argues that with the introduction of the landscape, research in string theory has become more like metaphysics than physics, and he suggests that string theorists have been less concerned with the lack of empirically testable consequences in recent years. We agree that some string theorists are very speculative, but it does not seem correct to say that string theorists in general have given up the hope of testability. But of course,  the longer it takes for string theorists to come up with anything testable, the more speculative the endeavour appears.

The recent suggestions that there really exists a multitude of universes, a landscape, seems wildly speculative. The critic might reasonably say that the only reason to believe in the real existence of a multitude of universes is that it explains, in a certain sense, the values of the constants in our universe. But this explanation only increases our wonder; in what lies the explanatory value of postulating a multitude of universes, enormously many more than the number of constants to be explained? It's hard to see any simplification or unification in this; what is needed are some independent arguments for the reality of the multiverse. It should also be noted that for there to be any kind of explanation it must be assumed that the other universes are actual and not mere possibilia. The idea of the `landscape' of universes has also been controversial within the physics community. 

A philosophically interesting conclusion to be drawn from the fact that string theory still dominates theoretical physics is that those involved in general have rather strong realist convictions. If they had no such convictions their interest and stubbornness in pursuing this research programme would be inexplicable.

Popular accounts of string theory, of which we have seen quite a number in recent years, have not always been sufficiently clear about its speculative character. In this respect it differs very much from for example GTR and the standard model.

As already indicated, we think Lakatos theory, MSRP, in Hacking's interpretation,  is the most viable view in methodology. But there is one aspect lacking in Lakatos account, viz., the possibility of a merger of differerent research programmes. We have seen some such examples in the history of science. One example is Schr\"{o}dinger's wave mechanics and Heisenberg's matrix mechanics.\footnote{The view that it was just found out that two formulations of quantum mechanics were equivalent is severely simplified. This has been argued in detail in Muller (1997).} Also string theory itself is an attempt to merge earlier research programmes.\footnote{See also the discussion on merging in Audretsch (1981), where he discusses quantum gravity in general.} The possibility of a similar merger of ideas from string theory and from some of its present competitors such as loop quantum gravity should be kept in mind. Different research programmes might be holding different parts of the puzzle. This we consider to be a good reason for arguing for a more open attitude between proponents of different research programmes especially when no programme is truly progressive at the empirical level.

The situation for those who decide about funding of research in theoretical physics is somewhat similar to investors in venture capital companies who want to invest money in projects that will be successful in the future and avoid waisting money on bad ideas. But no one knows in advance what will be successful and there are good reason to think that most projects will fail and few will succeed. So the venture capitalist had better not  put all eggs in the same basket, but to engage in several projects, in the hope that one or two of them will succeed and the profits in these will cover the lost money in the others. 

The parallel seems sufficiently close for a similar conclusion regarding funding of research; we had better not  use all money on one idea, but  to diversify. There are no strong reasons to think that string theory is the only way forward; another approach might very well be the correct one, and we cannot know in advance which.  

From our perspective we believe that a more pluralist approach is important in a situation where no research programme is empirically progressive. Still we think it is to be expected and even recommended that if one research programme starts to be empirically progressive this programme will receive most funds until it is again stuck and does not produce new empirical results. This we think is in general a reasonable strategy to use for people providing funds. 

For a new programme to become progressive there need to be risk takers that choose to work with a programme  with dim prospects.  To start a new research programme or early on join a new research programme is a high risk strategy. You will be hailed as a genius if you succeed and forgotten or seen as a crackpot if you fail, just as the risk taking entrepreneur that tries something new might become very rich or fail miserably. In both cases predictions are highly uncertain.

Just as it would not be good for the economy if we all were risk taking entrepreneurs it would not be good for the scientific community if all scientists were iconoclastic revolutionaries. A mix of risk takers, critics, and routine workers is needed. The scientific community needs different kinds of scientists working on different approaches and with different attitudes. Feynman expressed an analogous view some forty five years ago in his Nobel lecture:
\begin{quote}
Therefore, I think equation guessing might be the best method to proceed to obtain the laws for the part of physics which is presently unknown. Yet, when I was much younger, I tried this equation guessing and I have seen many students try this, but it is very easy to go off in wildly incorrect and impossible directions. I think the problem is not to find the best or most efficient method to proceed to a discovery, but to find any method at all. Physical reasoning does help some people to generate suggestions as to how the unknown may be related to the known.Theories of the known, which are described by different physical ideas may be equivalent in all their predictions and are hence scientifically indistinguishable. However, they are not psychologically identical when trying to move from that base into the unknown. For different views suggest different kinds of modifications which might be made and hence are not equivalent in the hypotheses one generates from them in ones attempt to understand what is not yet understood. I, therefore, think that a good theoretical physicist today might find it useful to have a wide range of physical viewpoints and mathematical expressions of the same theory (for example, of quantum electrodynamics) available to him. This may be asking too much of one man. Then new students should as a class have this. 

If every individual student follows the same current fashion in expressing and thinking about electrodynamics or field theory, then the variety of hypotheses being generated to understand strong interactions, say, is limited. Perhaps rightly so, for possibly the chance is high that the truth lies in the fashionable direction. But, on the off-chance that it is in another direction - a direction obvious from an unfashionable view of field theory - who will find it? Only someone who has sacrificed himself by teaching himself quantum electrodynamics from a peculiar and unusual point of view; one that he may have to invent for himself. I say sacrificed himself because he most likely will get nothing from it, because the truth may lie in another direction, perhaps even the fashionable one.\\
(Feynman (1965))
\end{quote}

\section*{Acknowledgement}

We thank Ulf Danielsson, Stanley Deser and two anonymous referees for valuable comments. The research behind this paper was supported by Riksbankens Jubileumsfond.

\section*{References}
\addcontentsline{toc}{section}{References}
Audretsch, J. (1981). Quantum gravity and the structure of scientific revolutions. \emph{Zeitschrift f\"{u}r allgemeine Wissenschaftstheorie}, XII/2.\\
Butterfield, J. \& Earman J. (eds.) (2007). \emph{Philosophy of physics: Part B}. North Holland.\\
Callender, C. \& Huggett, N. (eds.) (2001). \emph{Physics meets philosophy at the Planck scale}. Cambridge University Press.\\
Carnap, R. (1966). \emph{Philosophical foundations of physics: An introduction to the philosophy of science}. Basic Books.\\
Cartwright, N. \& Frigg R. (2007). String theory under scrutiny. \emph{Physics World}, September, pp. 14-15.\\
Chalmers, M (2007). Stringscape. \emph{Physics World}, September, pp. 35-47.\\
Davies, P.C.W. \& Brown J.R. (1988). \emph{Superstrings: A theory of everything?} Cambridge University Press.\\
Dawid, R. (2009). On the conflicting assessments of string theory. Philosophy of Science, 76(5).\\
Feynman, R.P. (1963). \emph{The Feynman Lectures on Physics}, Vol. 1. Addison-Wesley.\\
Feynman, R. P. (1965). \emph{Nobel lecture}. $\langle$ http://nobelprize.org/nobel$_{-}$prizes/\\physics/laureates/1965/feynman-lecture.html $\rangle$. \\
Feynman, R.P. (1992). \emph{The Character of Physical Law}. Penguin.(Originally published by BBC 1965 and based on his Messenger Lectures at Cornell 1964.)\\
Friedman, M. (1974). Explanation and scientific understanding. \emph{Journal of Philosophy 71}, 5-19.\\
Galison, P. (1995). Theory bound and unbound: Superstrings and experiment. In F. Weinert  (Ed.) \emph{Laws of Nature, essays on the, philosophical, scientific and historical dimensions.} Walter de Gruyter.\\
Green, M.B. \& Schwarz, J.H. (1984). Anomaly cancelations in supersymmetric D=10 gauge theory and superstring theory. \emph{Physics Letters B}, 149, 117\\ 
Green, M.B., Schwarz, J.H. \& Witten, E. (1987). \emph{Superstring theory: 2 volumes}. Cambridge University Press.\\
Greene, B. (1999). \emph{The elegant universe}. Jonathan Cape. \\
Greene, B. (2004). \emph{The fabric of the cosmos: Space, time and the texture of reality}. Allen Lane.\\
Gubser, S.S. (2009). Using string theory to study quark-gluon plasma: Progress and perils. \emph{Nuclear Physics A}, 830, 567c-664c.\\
Hacking, I. (1983). \emph{Representing and intervening}. Cambridge University Press.\\
Hedrich, R. (2006). String theory from physics to metaphysics. \emph{Physics and Philosophy}, ISSN 1863-7388-2006-ID:005\\
Hull, C.M. \& Townend, P. K. (1995). Unity of Superstring dualities. \emph{Nuclear Physics B}, 438, 109-137.\\
Kachru, S., Kalloch, R., Linde, A., \& Trivedi, S.P. (2003). de Sitter Vacua in String theory. \emph{Physical  Review D, 68}, 046005\\
Kitcher, P. (1981). Explanatory unification. \emph{Philosophy of Science}, 48, 507-531.\\
Kuhn, T. (1970). \emph{The structure of scientific revolutions} (2nd ed.). The University of Chicago Press.\\
Lakatos, I. (1970). Falsification and the methodology of scientific research programmes. In I.
Lakatos,  \& A. Musgrave (1970)(Eds.), \emph{Criticism and the Growth of Knowledge} (pp. 91-196). Cambridge University Press.\\
Lovelace, C. (1971). Pomeron form factors and dual Regge cuts. \emph{Physics Letters B}, 34, 500-506.\\
Mehra, J. (1994). \emph{The beat of a different drum: The life and science of Richard Feynman}. Oxford University Press.\\
Muller, F.A. (1997). The equivalence myth of quantum mechanics, published in two parts in \emph{Studies in History and Philosophy of Modern Physics}, 28, 35-61, 219-247, and an Addendum in 30 (1999) 543-545.\\
Nagel, E. (1961). \emph{The Structure of Science}. London: Rouledge \& Kegan Paul.\\
Nambu, Y. (1970). Quark model and the factorization of the Veneziano amplitude. In R.Chand (Ed.), \emph{Symmetries and quark models} (p. 269). Gordon and Breach.\\
Neveu, A., \& Schwarz, J.H. (1971). Factorizable dual model of pions. \emph{Nuclear Physics B, 31}, 529.\\
Nielsen, H.B. (1970). An almost physical interpretation of the integrand of the n-point Veneziano model. In \emph{15th International Conference on High Energy Physics}, Kiev, Unpublished.\\
Polchinski, J. (1995). Dirichlet-Branes and Ramond-Ramond Charges. \emph{Physics Review Letters}, 75, 4724.\\
Polchinski, J. (1998). \emph{String theory: 2 volumes}. Cambridge University Press.\\
Popper, K. (1959). \emph{The logic of scientific discovery}. Hutchinson.\\
Quine, W.V.O. (1969). \emph{Ontological relativity \& other essays}. Columbia University Press.\\
Ramond, P. (1971). Dual theory for free fermions, \emph{Physical Review D}, 3, 2415.\\ 
Rickles, D. (Ed.) (2008).\emph{The ashgate companion to contemporary philosophy of physics}. Ashgate Publishing Limited.\\
Rickles, D., French, S. \& Saatsi, J. (Eds.). (2006).\emph{The structural foundations of quantum gravity}. Oxford University Press.\\
Rovelli, C. (2007). Quantum Gravity. In J. Butterfield \&  J. Earman (Eds.), \emph{Philosophy of physics, part B} (pp. 1287-1330). North Holland.\\
Rugh, S.E. \& Zinkernagel, H. (2002). The quantum vacuum and the cosmological constant problem. \emph{Studies in History and Philosophy of Modern Physics}, 33, 663-705.\\
Scherk, J. \& Schwarz, J.H. (1974). Dual models for non-hadrons. \emph{Nuclear Physics B, 81}, 118.\\
Schnitzer, H.J. (2003). String Theory: A theory in search of an experiment. $\langle$arXiv:physics/0311047$\rangle$.\\
Smolin, L. (2006). \emph{The trouble with physics}. Houghton Mifflin Company.\\
Suppe, F. (Ed.).(1977). \emph{The structure of scientific theories}. University of Illinois Press.\\
Susskind, L. (1970). Structure of hadrons implied by duality.  \emph{Physical Review D, 1}, 1182.\\
Susskind, L. (2003). The anthropic landscape of string theory.$\langle$arXiv:hep-th/0302219v1$\rangle$.\\
Susskind, L. (2005). \emph{The cosmic landscape: String theory and the illusion of intelligent design}. Little, Brown \& Company.\\
Townsend, P.K., (1995) The eleven-dimensional supermembrane revisited. \emph{Physics Letters B} 350, 184.\\
Veneziano, G. (1968). Construction of a crossing symmetric, Regge-behaved amplitude for linearly rising trajectories. \emph{Nuovo Cimento}, 57A, 190.\\
Witten, E.  (1995), String theory dynamics in various dimensions. \emph{Nuclear Physics B} 443, 85.\\
Woit, P. (2006). \emph{Not Even Wrong}. Jonathan Cape.\\
Yoneya, T. (1973). Quantum gravity and the zero-slope limit of the generalized Virasoro model. \emph{Nuovo Cimento Letters}, 8, 951.\\
Zwiebach, B. (2004). \emph{A first course in string theory}. Cambridge University Press.\\

\end{document}